\documentclass[twocolumn,floatfix]{revtex4-2}
\pdfoutput=1
\usepackage{graphicx}
\usepackage{dcolumn}
\usepackage{longtable}
\usepackage{amsmath}
\usepackage{amssymb}

\begin{document}
\title{Signatures for shape coexistence and shape/phase transitions in even-even nuclei}

\author
{Dennis Bonatsos$^1$, Andriana Martinou$^1$,  S.K. Peroulis$^1$, T.J. Mertzimekis$^2$, and N. Minkov$^3$ }

\affiliation
{$^1$Institute of Nuclear and Particle Physics, National Centre for Scientific Research ``Demokritos'', GR-15310 Aghia Paraskevi, Attiki, Greece}

\affiliation
{$^2$  Department of Physics, National and Kapodistrian University of Athens, Zografou Campus, GR-15784 Athens, Greece}

\affiliation
{$^3$Institute of Nuclear Research and Nuclear Energy, Bulgarian Academy of Sciences, 72 Tzarigrad Road, 1784 Sofia, Bulgaria}

\begin{abstract}

Systematics of B(E2) transition rates connecting the first excited $0_2^+$ state to the first excited $2_1^+$ state of the ground state band in even-even nuclei indicates that shape coexistence of the ground state band and the first excited $K=0$ band should be expected in nuclei lying within the stripes of nucleon numbers 7-8, 17-20, 34-40, 59-70, 96-112 predicted by the dual shell mechanism of the proxy-SU(3) model, avoiding their junctions, within which high deformation is expected. Systematics of the excitation energies of the $0_2^+$ states in even-even nuclei show that shape coexistence due to proton-induced neutron particle-hole excitations is related to a first-order shape/phase transition from spherical to deformed shapes, while shape coexistence due to neutron-induced proton particle-hole excitations is observed along major proton shell closures.

 \end{abstract}

\maketitle
\section{Introduction}   

Shape coexistence (SC) \cite{Meyer,Wood,HW,Garrett} is known to occur in several nuclei across the nuclear chart, in which the ground state band and an excited $K=0$ band lying close in energy exhibit substantially different structures, for example with one of them being spherical and the other being deformed. The advent of radioactive ion beam facilities around the world makes possible the search for SC in regions yet unexplored. The question then arises, where to look for new examples of SC.

While the initial expectation was that SC can occur anywhere across the nuclear chart, already in 2011, in Fig. 8 of the authoritative review of Ref. \cite{HW}, it has been observed that nuclei exhibiting experimental signs of SC appear to cluster together into certain islands on the nuclear chart. A possible theoretical justification of the origin of these islands has been given recently in the framework of the proxy-SU(3) symmetry \cite{proxy1,proxy2,EPJASM}, which restores the SU(3) symmetry of the 3-dimensional isotropic harmonic oscillator (3D-HO) in the nuclear shells beyond the $sd$ shell through replacement of the intruder orbitals by their ``proxies'' (which have escaped to the shell below because of the action of the spin-orbit interaction), based on the similarities of the Nilsson quantum numbers characterizing the two sets of orbitals (see the recent review \cite{Symm2} for further details). Within this framework, a dual shell mechanism, based on the interplay of the magic numbers of the 3D-HO and the familiar nuclear magic numbers created by the spin-orbit interaction, has been suggested in 2021 \cite{EPJASC}. According to the dual shell mechanism,  SC can occur only within certain stripes on the nuclear chart, occurring within the nucleon numbers 7-8, 17-20, 34-40, 59-70, 96-112. However, this is only a necessary condition and not a sufficient one.

Subsequent searches for a sufficient condition for the occurrence of SC have been based on the assumptions about its microscopic origin. It has been accepted over the years that SC occurs when particle-hole (p-h) excitations across major shells are possible \cite{Meyer,Wood,HW,Garrett}, the regions around $Z=82$ and $Z=50$ providing the most robust examples. Calculations of single-particle energy spectra using covariant density functional theory have recently corroborated this assumption, indicating that proton p-h excitations leading to SC occur in certain islands of the nuclear chart, centered around the $Z=82$ and $Z=50$ major shell gaps \cite{CDFTPLB,CDFTPRC}.
This has been called neutron-induced SC, since they are caused by the varying in number neutrons acting on the same set of protons.  Furthermore, it has been suggested that islands of SC occurring around  $N=90$ and $N=60$ can be attributed to neutron p-h excitations across the $N=112$ and $N=70$ shell closures of the three-dimensional isotropic harmonic  oscillator (3D-HO), signifying the beginning of the participation of the intruder (opposite parity) orbitals $1i_{13/2}$ and $1h_{11/2}$ to the onset of nuclear deformation \cite{CDFTPLB,CDFTPRC}. This has been called proton-induced SC, since they are caused by the varying in number protons acting on the same set of neutrons. In addition, a region around $N=40$, $Z=40$ has been found, in which both proton p-h excitations and neutron p-h excitations are possible \cite{CDFTPRC}.

It is the purpose of the present work to show that simple signatures of SC exist in even-even nuclei, based on the energies of the first excited $L=0$ and $L=2$ states, $0_2^+$ and $2_1^+$ respectively, and the transition rates $B(E2; 0_2^+\to 2_1^+)$ and $B(E2; 2_1^+ \to 0_1^+)$, involving these states and the ground state, $0_1^+$. These simple signatures not only cut off large parts of the SC stripes determined as a  necessary condition by the proxy-SU(3) approach, but in addition clarify the connection between SC and shape/phase transitions in the $N=90$ and $N=60$ regions, an issue having attracted considerable amount of recent work \cite{Fossion,Ramos100,Ramos102,Ramos105}. 

\section{New signatures for shape coexistence}

In Table I \textsl{all} nuclei beyond $Z=18$, $N=18$ for which a $K=0$ band is experimentally known \cite{ENSDF} and the transition rate $B(E2; 0_2^+\to 2_1^+)$ is also known \cite{ENSDF} are collected, along with the involved experimental energies and B(E2) transition rates \cite{ENSDF}. The energy ratio 
\begin{equation}\label{R42}
R_{4/2}= \frac{E(4_1^+)}{E(2_1^+)}, 
\end{equation}
of the first two excited states of the ground state band is also shown, since it is a well known and easily measurable indicator of collectivity, with deformed nuclei having $R_{4/2}>3$, transitional nuclei exhibiting $2.4 < R_{4/2} < 3$, and vibrational nuclei possessing $R_{4/2}<2.4$~. In addition, the ratio of transition rates 
\begin{equation}\label{B02}
B_{02} =  \frac{B(E2; 0_2^+\to 2_1^+)}{B(E2; 2_1^+ \to 0_1^+)}
\end{equation}  
and  the energy ratio 
\begin{equation}\label{R02}
R_{0/2}= \frac{E(0_2^+)}{E(2_1^+)} 
\end{equation}   
are listed. The inverse ratio 
\begin{equation}\label{R20} 
R_{2/0}=1/R_{0/2} 
\end{equation}
will also be used later.  Deformed nuclei with $R_{4/2} > 3.05$ are shown in boldface. For the $N=90$ isotones $^{150}$Nd, $^{152}$Sm, and $^{154}$Gd, which are known to be the best experimental manifestations \cite{McCutchan,CastenPPNP,Cejnar} of the X(5) critical point symmetry \cite{IacX5} from spherical to prolate deformed nuclei, the $R_{4/2}$, $R_{0/2}$ and $B_{02}$ ratios are shown in boldface. The same is done for the $N=60$ isotones $^{98}$Sr, 
$^{100}$Zr, $^{102}$Mo, which have been characterized as lying at the transition point from spherical to deformed shapes \cite{Ramos102,Ramos105,Esmaylzadeh104}. The following observations can be made. 

1) The $N=90$ isotones $^{150}$Nd, $^{152}$Sm, and $^{154}$Gd, have experimental $R_{4/2}$ and $R_{0/2}$ values close to the X(5) ones, while their $B_{02}$ experimental values are lower than the X(5) value by a factor of two. 

2) In contrast, the $N=60$ isotones  $^{98}$Sr, $^{100}$Zr, and $^{102}$Mo, have $B_{02}$ values higher than the X(5) one, but $R_{0/2}$ values lower than the X(5) one by a factor of 3, while their $R_{4/2}$ values are also lower than the X(5) one (except for $^{98}$Sr). 

The $B_{02}$ ratio is plotted vs. the $R_{0/2}$ ratio in Figure 1(a). High values of $B_{02}$ are observed at low $R_{0/2}$, while at large values of $R_{0/2}$ the ratio $B_{02}$ collapses to very low values, the two regions being separated by a bump, formed by the $N=90$ isotones $^{150}$Nd, $^{152}$Sm, and $^{154}$Gd. 

Combining Table I and Fig. 1(a) we conclude that the nuclei of Table I can be divided into two groups. The first group comprises  nuclei lying in Fig. 1(a) before the X(5) bump, in general characterized by $R_{4/2}<3.05$, $R_{02}<5.7$, and $B_{02}>0.1$, while the second group contains nuclei lying in Fig. 1(a) after the X(5) bump, in general characterized by $R_{4/2}>3.05$, $R_{02}>5.7$, and $B_{02}<0.1$. It is worth remarking that the nuclei of the first group have been reported as exhibiting SC (see Table I for relevant references), while SC is absent from the nuclei of the second group.  

In corroboration of these conclusions, the parameter-independent predictions of various special solutions of the Bohr Hamiltonian \cite{Bohr} are shown in Fig. 1(b). These include X(5) \cite{IacX5}, which uses an infinite square well potential in the collective variable $\beta$, its modifications X(5)-$\beta^{2n}$, $n=1,2,3,4$ \cite{BonX5}, in which the $\beta^{2n}$, $n=1,2,3,4$  potentials are used instead of the infinite square well potential, and the  confined $\beta$-soft (CBS) solution \cite{Gorbachenko}, which uses an infinite square well potential displaced from the origin, having boundaries at $\beta_M > \beta_m \geq 0$ and characterized by the ratio $r_\beta=\beta_m/\beta_M$. It is clear that as the potential becomes more narrow, first by gradually going from the harmonic oscillator shape (X(5)-$\beta^2$) to the the infinite square well (X(5)), and then by moving the left wall of the infinite square well towards the right wall (CBS), $R_{0/2}$ raises and $B_{02}$ drops.   
The same qualitative behavior seen in Fig. 1(a) is observed also in Fig. 1(b).

The qualitative message conveyed by Fig. 1 is that in well deformed nuclei the $K=0$ band raises high in energy, thus making its connection through B(E2) transition rates to the ground state weak. The large energy difference between the two bands excludes any possiblity for SC in well deformed nuclei.  

It is instructive at this point to consider the $P$-factor \cite{Brenner}
\begin{equation}\label{P}
P= \frac{N_p N_n}{N_p+N_n}, 
\end{equation}
where $N_p$ ($N_n$) is the number of valence protons (neutrons) measured from the nearest closed shell. The $P$-factor is a measure of collectivity, expressing the relative strength of the quadrupole-quadrupole interaction over the strength of the pairing interaction. It is known \cite{Brenner,McCutchan,CastenPPNP} that deformed nuclei are characterized by $P>5$, while nearly spherical and transitional nuclei have $P<5$. Nuclei with $P \approx 5$ are known to form contours  \cite{McCutchan,CastenPPNP} in the nuclear chart, also shown in Fig. 2 as orange boxes.  The contours are drawn by calculating the $P$-factor for all nuclei in the area under consideration and selecting the values closest to $P=5$ \cite{McCutchan,CastenPPNP}. 

We now place the nuclei of Table I on the nuclear chart in Fig. 2, representing nuclei of Table I with $R_{4/2}<3.05$ ($R_{4/2}>3.05$) by green (blue) full triangles. 
Deformed nuclei with $P>5$ lie inside the $P\approx 5$ contours \cite{McCutchan,CastenPPNP}, while  nearly spherical and transitional nuclei with $P<5$ lie outside them. 

In Fig. 2 we observe that the $N=90$ isotones, which are the best experimental manifestations \cite{McCutchan,CastenPPNP} of the X(5) critical point symmetry, lie on the relevant $P\approx 5$ contour  or are touching it. Exactly the same situation is seen around $Z=40$, $N=60$, with $^{98}$Sr and $^{100}$Zr falling on the $P\approx 5$  
 contour, and with  $^{96}$Sr and $^{98}$Zr touching it. Therefore Fig. 2 suggests that a shape/phase transition is occurring in the region of $Z=40$ and $N=60$, similar to the one seen in the region of $Z=60$ and $N=90$, as already suggested in Refs. \cite{Ramos100,Ramos102,Ramos105}.    

In Fig. 2 the stripes  corresponding to 7-8, 17-20, 34-40, 59-70, 96-112, 146-168 protons or neutrons, within which SC can be expected to occur according to the dual shell mechanism \cite{EPJASC} developed within the proxy-SU(3) model  \cite{proxy1,proxy2,EPJASM}  are also shown. We see in general that nuclei exhibiting SC are lying within the proxy-SU(3) stripes and outside the $P\approx 5$ contours, while deformed nuclei exhibiting no SC are falling within the $P\approx 5$ contours . In other words, the $P\approx 5$ contours  cut off large parts of the proxy-SU(3) stripes at their junctions, preventing the appearance of SC within them. 

Figs. 1 and 2  imply that SC is expected to occur in nuclei which are not well deformed (i.e., $R_{4/2}<3.05$), having strong B(E2)s (i.e., $B_{02}>0.1$) connecting the $0_2^+$ and $2_1^+$ states, which are lying quite close to each other in energy  (i.e., $R_{0/2}< 5.7$). 

In corroboration of these findings, we have collected in Table II, which is similar to Table I, \textsl{all} nuclei beyond $Z=18$, $N=18$ for which the transition rate $B(E2; 0_2^+\to 2_1^+)$ is known \cite{ENSDF}, but no other levels of a $K=0$ band based on the $0_2^+$ state are known. Adding these nuclei in Fig. 2, and representing nuclei in which SC does (does not) appear by green (blue) full circles,  we see in general that nuclei exhibiting SC are lying within the proxy-SU(3) stripes and outside the $P\approx 5$ contours, while nuclei exhibiting no SC occur outside the stripes predicted by the proxy-SU(3) dual shell mechanism. 

In summary, in Fig. 2 it becomes evident that nuclei exhibiting SC (shown in green color) fall within the proxy-SU(3) stripes and outside the $P\approx 5$ contours, while deformed nuclei showing no SC (shown as blue full triangles) fall within the $P\approx 5$ contours  and non-deformed nuclei showing no SC (shown as blue full circles) fall outside the proxy-SU(3) stripes. Fig. 2 suggests that when searching for SC one should avoid well deformed nuclei with $R_{4/2}> 3.05$, and among nuclei with $R_{4/2} < 3.05$ one should focus attention on nuclei falling within the stripes predicted by the proxy-SU(3) dual shell mechanism, exhibiting $R_{0/2}<5.7$ and $B_{02}>0.1$. It should be noticed that this prescription appears to hold for nuclei which would exhibit SC of their ground state band with a nearby excited $K=0$ band. It does not exclude the possibility of SC exhibited by two excited bands within the same nucleus.  

In Fig. 2 an exception from the above mentioned systematics is observed for the Pt isotopes, which  appear outside the SC stripes predicted by the dual shell mechanism of proxy-SU(3). It has been shown that the shape evolution in the Pt isotopes can be described \cite{McCutchan71} in the framework of the interacting boson model (IBM) without using particle-hole excitations. 

It should be noticed that from Fig. 2 they are absent the neutron-deficient Hg, Pb, and Po isotopes, which form the region in which SC was first discovered  
and interpreted in terms of two-particle--two-hole excitations across the $Z=82$ energy gap \cite{Wood,HW,Garrett}. The reason behind this absence is that no data for   
$B(E2; 0_2^+\to 2_1^+)$ in these nuclei is found in the relevant database \cite{ENSDF}. In order to cope with this deficiency, we have collected in Table III all Hg, Pb, and Po isotopes for which SC has been suggested and for which at least the $0_2^+$ state is experimentally known according to Ref. \cite{ENSDF}. Placing these nuclei in Fig. 2 (representing them as purple diamonds), we see that all Hg and Pb isotopes lie within the $N=96$-112 stripe, in which SC is predicted to occur by the dual shell mechanism, while the Po isotopes show some tendency to extend beyond the $N=112$ border. In Table III we see that the conditions $R_{0/2}<5.7$ and $R_{4/2}<3.05$ are fulfilled in all cases (except for $R_{4/2}$ in $^{182}$Hg, which comes from an uncertain assignment of a 1124.8 keV state as the $4_1^+$ member of the ground state band \cite{ENSDF}). It would therefore be very interesting to measure the $B(E2; 0_2^+\to 2_1^+)$ transition rate in these nuclei, in order to test if they satisfy the condition $B_{02}>0.1$ as well. In an extensive search made in the literature (see Table III for the relevant references), only theoretical predictions for $^{186,188}$Pb \cite{Nomura2012b} and $^{198}$Po \cite{GarciaRamos2015} have been found, the latter accompanied by experimental data as well. These findings are reported in Table IV. In all cases, the  $B_{02}>0.1$ condition is well satisfied.  

\section{Relation of shape coexistence to shape/phase transitions}  

It is worth looking at the behavior of the ratios $R_{2/0}$ and $R_{4/2}$ when the shape/phase transition from spherical to deformed shapes is taking place. This is depicted within the interacting boson model (IBM) \cite{IA} in Fig. 3, using the extended consistent $Q$ formalism \cite{Warner,Lipas} of the IBM with a Hamiltonian \cite{Werner} 
\begin{equation} \label{H}
H(\zeta,\chi) = c \left[ (1-\zeta) \hat n_d -{\zeta\over 4 N_B}
\hat Q^\chi \cdot \hat Q^\chi\right],
\end{equation}
\noindent where pairing is represented by the term $\hat n_d = d^\dagger \cdot \tilde d$, the quadrupole operator is 
$\hat
Q^\chi = (s^\dagger \tilde d + d^\dagger s) +\chi (d^\dagger\tilde d)^{(2)},$ $N_B$ is the number of valence bosons, and $c$
is a scaling factor. The above Hamiltonian contains two
parameters, $\zeta$ and $\chi$. The parameter $\zeta$ extends  from 0 to 1, while the parameter 
 $\chi$ is ranging from 0 to $-\sqrt{7}/2$. In this
parameterization, the three dynamical symmetries of the IBM \cite{IA} are given by
$\zeta$ = 0, any $\chi$ for U(5), $\zeta$ = 1, $\chi$ =
$-\sqrt{7}/2$ for SU(3) and $\zeta$ = 1.0, $\chi$ = 0.0 for O(6).
Calculations were performed with the code IBAR \cite{IBAR}, which
allows boson numbers up to 1000. For $\chi=-\sqrt{7}/2$ the critical point is known to be given by $\zeta_{crit} = 16N_B/ (34 N_B-27)$ \cite{Fernandes}, which in the case of $N_B=250$ gives $\zeta_{crit}= 0.4721$. Indeed in Fig. 3 we see that in the region of  $\zeta_{crit}$ the ratio $R_{2/0}$ first jumps from the vibrational value 0.5 to a high value and then collapses to rotational very small values, while the ratio $R_{4/2}$, first sinks from the vibrational value 2.0 to a low value and then it rises abruptly to the rotational value 3.33~. A large number of bosons ($N_B=250$) has been used in order to exhibit the changes more clearly, since these are moderated at realistic boson numbers, as seen for example in Ref. \cite{order}, although the qualitative behavior remains the same.  

In the above, the ratio $R_{2/0}$ serves as the order parameter, while the parameter $\zeta$ serves as the control parameter of the first order shape/phase transition (SPT) from spherical to prolate deformed shapes \cite{IacX5,McCutchan,CastenPPNP,Cejnar,Deans,order}.   

The experimental $R_{2/0}$ ratios for nuclei in the $Z=60$, $N=90$ and $Z=40$, $N=60$ regions are shown in Figs. 4(a), (b). In this case, the neutron number $N$ is used as the control parameter of the SPT \cite{McCutchan,CastenPPNP,Cejnar,order}. It is clear that a behavior similar to that of Fig. 3(a) is seen in Fig. 4(a) for the Nd, Sm, Gd, and Dy series of isotopes, fading away for the Er isotopes, while in Fig. 4(b) this behavior is seen more clearly in the Zr isotopes, as well as in the Mo and Sr isotopes to a lesser extend, fading away in the Ru and Pd isotopes. These observations are in agreement to studies connecting the appearance of SC in these regions to the shape/phase transition from spherical to deformed shapes \cite{Ramos100,Ramos102,Ramos105}. In Fig. 4(c) a similar behavior is seen at the $N=40$ subshell closure for the Ni, Zn, Ge, Se isotopes. 

The critical behavior seen in Fig. 4 can be associated with the beginning of participation of the intruder orbitals to the onset of deformation \cite{FP1,FP2,FP3}. In particular  

a) $N=90$ signifies the beginning of participation of the intruder $1i_{13/2}$ orbital. As seen in Refs. \cite{CDFTPLB,CDFTPRC} in the framework of covariant density functional theory (CDFT), particle-hole excitations occur in which the orbitals 1/2[660] or  3/2[651] of $1i_{13/2}$ get occupied, while the orbital 5/2[523] of $2f_{7/2}$ becomes vacant. 
     
b) $N=60$ signifies the beginning of participation of the intruder $1h_{11/2}$ orbital. As seen in Refs. \cite{CDFTPLB,CDFTPRC} in the framework of CDFT, particle-hole excitations occur in which the orbitals 1/2[550] or  3/2[541] of $1h_{11/2}$ get occupied, while the orbitals 5/2[413] of $2d_{5/2}$ or 1/2[411] of $3s_{1/2}$  become vacant. The crossing of regular and intruder configurations in this region has also been suggested in the framework of the interacting boson model with configution mixing (IBM-CM) \cite{Ramos100,Ramos102,Ramos105}.  

c) $N=40$ signifies the beginning of participation of the intruder $1g_{9/2}$ orbital.  As seen in Refs. \cite{CDFTPLB,CDFTPRC} in the framework of CDFT, particle-hole excitations occur in which the orbitals 1/2[440], and/or  3/2[431], and/or 5/2[422] of $1g_{9/2}$ get occupied, while the orbitals 5/2[303] and/or 3/2[301] of $1f_{5/2}$ and/or 1/2[301] of $2p_{1/2}$  become vacant. 

The present findings further emphasize the difference between SC seen in the regions around $Z=82$, 50 and in the regions around $N=90$, 60. In the first set SC is due to proton particle-hole (p-h) excitations caused by the neutrons (neutron-induced SC), while in the second set SC is due to neutron particle-hole excitations caused by the protons  (proton-induced SC). In the first case,  p-h excitations appear across major proton shell closures, while in the second case, p-h excitations appear across neutron shells of different parity. In the first case a change of major shells occurs, while in the second case a shape/phase transition from spherical to prolate deformed shapes occurs. The first case corresponds to the islands L and J in Fig. 8 of Ref. \cite{HW}, while the second case corresponds to islands K and I in the same figure. In the $Z=40$, $N=40$ region, both mechanisms appear, corresponding to island I in Fig. 8 of Ref. \cite{HW}. 

\section{Conclusions}

The main findings of the present work are summarized here.

a) A practical guide for the experimental search of shape coexistence of the ground state band with an excited $K=0$ band is obtained, including the following steps: 

$\bullet$ SC should be searched for within the stripes occurring between  7-8, 17-20, 34-40, 59-70, 96-112 nucleons, predicted by the dual shell mechanism developed within the proxy-SU(3) model, but avoiding the $P\approx 5$ bubbles appearing at the junctions of these stripes, within which highly deformed nuclei appear. 

$\bullet$ Candidates for SC should possess $R_{4/2} < 3.05$, $R_{0/2} < 5.7$, $B_{02}>0.1$. Qualitatively these conditions mean that the nucleus should not be well deformed, its $K=0$ band has to lie close in energy to the ground state band, and the $K=0$ band-head should be strongly connected by a B(E2) transition rate to the first excited $L=2$ state of the ground state band.  

b) Two different mechanisms  leading to SC are identified:

$\bullet$ Proton particle-hole excitations across the major proton shell closures at $Z=82$, 50.

$\bullet$ Neutron particle-hole excitations at $N=90$, 60, 40, where intruder orbitals start participating to the onset of deformation, while a shape/phase transition from spherical to prolate deformed shapes is taking place in parallel. 

 It would be interesting to extend these considerations to nuclei with lower masses, in which an elongated island along the $N=Z$ line also appears, labeled as island A in Fig. 8 of Ref. \cite{HW}. Furthermore, it is well known  \cite{Meyer,Wood,HW,Garrett} that the Hg, Pb, and Po series of isotopes provide some of the best examples of shape coexistence throughout the nuclear chart. As emphasized through Tables III and IV, further experiments focusing on the measurement of the $B(E2; 0_2^+\to 2_1^+)$ transition rate in these nuclei would be interesting to pursue.

\section*{Acknowledgements} 

Support by the Bulgarian National Science Fund (BNSF) under Contract No. KP-06-N48/1  is gratefully acknowledged.


\begin{table*}

\caption{Nuclei beyond $Z=18$, $N=18$ with experimentally known \cite{ENSDF} $K=0$ bands and $B(E2;0_2^+\to 2_1^+)$ transition rates. Energies are given in keV, while $B(E2)$s are given in W.u.~. All data have been taken from Ref. \cite{ENSDF}. Deformed nuclei with $R_{4/2} > 3.05$ are shown in boldface. For $N=90$ isotones, which are good examples \cite{McCutchan,CastenPPNP,Cejnar} of the X(5) critical point symmetry \cite{IacX5}, the $R_{4/2}$, $R_{0/2}$ and $B_{02}$ ratios are shown in boldface. The same is done for the $N=60$ isotones which have been characterized \cite{Ramos100,Ramos102,Ramos105} as lying at the transition point from spherical to deformed shapes. The parameter-independent values \cite{IacX5,BonX5} provided by X(5) for the various ratios are reported in the last line of the table for comparison. References discussing the presence of SC in the particular nucleus are given in the last column. See Sec. II for further discussion. }

\bigskip

\begin{tabular}{ r r r r c c r r l  }

\hline
nuc. & $R_{4/2}$ & $E(2_1^+)$ & $E(0_2^+)$ & $B(E2; 2_1^+\to 0_1^+)$ & $B(E2; 0_2^+\to 2_1^+)$ & $R_{0/2}$ & $B_{02}$ & Refs. \\
 \hline

$^{40}$Ar & 1.980 & 1460.8 & 2120.9 & 9.0 (4)        & 5.3 (8)        & 1.452 & 0.589 & \cite{Wood,HW,Garrett} \\
$^{40}$Ca &       & 5629.4 & 3352.6 & 0.143 (+35-24) & 7.5 (+30-20)   & 0.639 &52.558 & \cite{Wood,HW,Garrett} \\ 
$^{42}$Ca & 1.805 & 1524.7 & 1837.3 & 9.5(4)         & 55. (5)        & 1.205 & 5.789 & \cite{Wood,HW,Garrett} \\
$^{70}$Ge & 2.071 & 1039.5 & 1215.6 & 20.8 (4)       & 48. (7)        & 1.169 & 2.308 & \cite{HW,Garrett,Guo76} \\
$^{72}$Ge & 2.072 &  834.0 &  691.4 & 23.5 (4)       & 89.0 (15)      & 0.829 & 3.787 & \cite{HW,Garrett,Guo76} \\
$^{72}$Se & 1.899 &  862.1 &  937.2 & 23.7 (17)      & 162. (28)      & 1.087 & 6.835 & \cite{Wood,HW,Garrett,McCutchan83,Mukherjee105,Budaca990} \\
$^{96}$Sr & 2.000 &  814.9 & 1229.3 & 13. (8)        & 15.3 (16)      & 1.509 & 1.177 & 
 \cite{HW,Garrett,Petrovici85,Ramos105,Clement116,Xiang873,Nomura94} \\
$^{98}$Sr & \textbf{3.006} &  144.2 &  215.6 & 96. (3)  & 62. (+7-6)  & \textbf{1.495} & \textbf{0.646}  & 
\cite{Wood,HW,Garrett,Park93,Ramos105,Clement116,Xiang873,Nomura94}\\
$^{98}$Zr & 1.674 & 1222.9 &  854.0 & 2.9 (+8-5)     & 145. (+40-30)  & 0.698 &50.000  & 
\cite{Wood,HW,Garrett,CDFTPRC,Petrovici85,Witt98,Ramos100,Ramos102,Togashi117,Nomura94,Gavrielov99,Gavrielov105}\\
$^{100}$Zr& \textbf{2.656} &  212.5 &  331.1 & 77. (2)  & 67. (6) & \textbf{1.558} & \textbf{0.870}  &
 \cite{Wood,HW,Garrett,CDFTPRC,Ramos100,Ramos102,Togashi117,Nomura94,Gavrielov99,Gavrielov105}\\

$^{100}$Mo& 2.121 &  535.6 &  695.1 & 37.6 (9)       & 89. (3)        & 1.298 & 2.367 & \cite{HW,Garrett,Abusara95,Nomura94} \\
$^{102}$Mo& \textbf{2.507} &  296.6 &  698.3 & 74. (9)        & 70. (30)       & \textbf{2.354} & \textbf{0.946} & 
 \cite{Wood,HW,Garrett,Abusara95,Esmaylzadeh104,Nomura94}\\ 
$^{104}$Ru& 2.482 &  358.0 &  988.3 & 57.9 (11)      & 25. (3)        & 2.761 & 0.432 &  \cite{HW,Garrett,Abusara95,Nomura94}\\
$^{110}$Pd& 2.463 &  373.8 &  946.7 & 55.5 (9)       & 37. (4)        & 2.533 & 0.667 & \cite{Wood,HW,Garrett}\\
$^{112}$Cd& 2.292 &  617.5 & 1224.3 & 30.31 (19)     & 51. (14)       & 1.983 & 1.683 & \cite{Wood,Garrett,Garrett101,Nomura106,Garrett123}\\
$^{116}$Cd& 2.375 &  513.5 & 1380.3 & 33.5 (12)      & 30. (6)        & 2.688 & 0.896 & \cite{Garrett,Nomura106}\\
$^{114}$Sn& 1.683 & 1299.9 & 1953.3 & 15. (3)        & 22. (8)        & 1.503 & 1.467  & \cite{Wood,Garrett,Spieker97}\\
$^{116}$Sn& 1.848 & 1293.6 & 1756.9 & 12.4 (4)       & 18. (3)        & 1.358 & 1.452 & \cite{Wood,Garrett,Petrache99}\\
$^{118}$Sn& 1.854 & 1229.7 & 1758.3 & 12.1 (15)      & 19. (3)        & 1.430 & 1.570 & \cite{Wood,Garrett}\\
$^{126}$Xe& 2.424 &  388.6 & 1313.9 & 44. (4)        & 6.4 (12)       & 3.381 & 0.145 & \\  
   
$^{148}$Nd& 2.493 &  301.7 &  916.9 & 57.9 (22)      & 3.12 (22)      & 3.039 & 0.539 &  \\
$^{150}$Nd& \textbf{2.927} &  130.2 &  675.9 & 116. (3)       & 43.1 (23)      & \textbf{5.191} & \textbf{0.372} & \cite{CDFTPRC} \\
$^{152}$Sm& \textbf{3.009} &  121.8 &  684.8 & 145.0 (16)     & 33.3 (12)      & \textbf{5.622} & \textbf{0.230} & \cite{HW,CDFTPRC,Garrett103}\\
$^{154}$\textbf{Sm}& 3.255 &   82.0 & 1099.3 & 176. (1)       & 12. (3)        &13.406 & 0.068 & \cite{CDFTPRC}\\
$^{152}$Gd& 2.194 &  344.3 &  615.4 & 73 (+7-6)      & 178. (+33-53)  & 1.787 & 2.438 &  \\ 
$^{154}$Gd& \textbf{3.015} &  123.1 &  680.7 & 157. 1         & 52. (8)        & \textbf{5.530} & \textbf{0.331} & \cite{HW,CDFTPRC}\\
$^{156}$\textbf{Gd}& 3.239 &   89.0 & 1049.5 & 189. 3         & 8. (+4-7)      &11.792 & 0.042 & \cite{CDFTPRC} \\
$^{158}$\textbf{Gd}& 3.288 &   79.5 & 1196.2 & 198. 5         & 1.17 (+418-13) &15.047 & 0.006 &  \\
$^{166}$\textbf{Er}& 3.289 &   80.6 & 1460.0 & 217. 5         & 2.7 (10)       &18.114 & 0.012 \\
$^{172}$\textbf{Yb}& 3.305 &   78.7 & 1042.9 & 212. 2         & 3.6 (10)       &13.252 & 0.017 \\

$^{174}$\textbf{Yb}& 3.310 &   76.5 & 1487.1 & 201. (7)       & 1.4 (+11-5)    &19.439 & 0.007 \\ 
$^{186}$\textbf{Os}& 3.165 &  137.2 & 1061.0 & 93.6 (21)      & 0.066          & 7.733 & 0.001 & \cite{Nomura84}\\
$^{192}$Os& 2.820 &  205.8 &  956.5 & 62.1 (7)       & 0.57 (12)      & 4.648 & 0.009 & \cite{Nomura84}\\
$^{196}$Pt& 2.465 &  355.7 & 1135.3 & 40.60 (20)     & 2.8 (15)       & 3.192 & 0.069 & \cite{Wood,Garrett,CDFTPRC,McCutchan71,Nomura83,Nomura84}\\

\hline

X(5)     & 2.904 &  & & & & 5.649 & 0.624 \\

\hline

\end{tabular}
\end{table*}

\begin{table*}

\caption{Nuclei beyond $Z=18$, $N=18$ in which the $B(E2;0_2^+\to 2_1^+)$ transition rates are experimentally known \cite{ENSDF} , 
but no other levels of a $K=0$ band based on the $0_2^+$ state are known. Energies are given in keV, while $B(E2)$s are given in W.u.~. All data have been taken from Ref. \cite{ENSDF}. References discussing the presence or absence of SC in the particular nucleus are given in the last column. Nuclei for which the absence of SC is expected or predicted are shown in boldface.  See Sec. II for further discussion. }

\bigskip

\begin{tabular}{ r r r r c c r r l  }

\hline
nuc. & $R_{4/2}$ & $E(2_1^+)$ & $E(0_2^+)$ & $B(E2; 2_1^+\to 0_1^+)$ & $B(E2; 0_2^+\to 2_1^+)$ & $R_{0/2}$ & $B_{02}$ & Refs. \\
 \hline

$^{38}$Ar &       & 2167.5 & 3376.9 & 3.40 (16) & 1.26 (8)          & 1.558 & 0.371 & \cite{Wood,HW,Garrett} \\
$^{44}$Ca & 1.973 & 1157.0 & 1883.5 & 10.9 (6)  & 22. (7)           & 1.628 & 2.018 & \cite{Wood,HW,Garrett} \\
$^{48}$Ca & 1.175 & 3831.7 & 4283.3 & 1.84 (+17-14) & 10.1 (5)      & 1.118 & 5.489 & \cite{Wood} \\
$^{46}$\textbf{Ti} & 2.260 &  889.3 & 2611.0 & 19.5 (6) & 50. (14)           & 2.936 & 2.564 & \\
$^{48}$\textbf{Ti} & 2.344 &  983.5 & 2997.2 & 13.2 (+13-11) & 20.6 (+44-32) & 3.047 & 1.561 & \\
$^{50}$\textbf{Ti} & 1.722 & 1553.8 & 3868.3 & 5.46 (19) & 1.6 (+14-5)       & 2.490 & 0.293 & \\
$^{54}$\textbf{Cr} & 2.185 &  834.9 & 2829.6 & 14.4 (6) &  10. (+3-4)       & 3.389 & 0.694 & \\
$^{56}$\textbf{Fe} & 2.462 &  846.8 & 2941.5 & 16.8 (7) & 2.4 (+7-12)        & 3.474 & 0.143 & \\

$^{58}$Ni & 1.691 & 1454.2 & 2942.6 & 10.0 (4) & 0.00040 (6)    & 2.023 & 4 $10^{-5}$ & \cite{Garrett} \\
$^{64}$Ni & 1.940 & 1345.8 & 2867.3 & 7.76 (26) & 3.15 (+23-21) & 2.131 & 0.406 & \cite{Garrett} \\ 
$^{64}$Zn & 2.326 &  991.6 & 1910.3 & 20.0 (5) & 0.057 (3)      & 1.927 & 0.003 & \cite{Garrett} \\
$^{68}$Zn & 2.244 & 1077.4 & 1655.9 & 14.69 (19) & 5.5 (10)     & 1.537 & 0.374 & \cite{Garrett,HW} \\
$^{70}$Zn & 2.019 &  884.9 & 1070.8 & 16.7 (10) & 37.3 (19)     & 1.210 & 2.234 & \cite{Garrett} \\

$^{74}$Ge & 2.457 &  595.9 & 1482.8 & 33.0 (4) & 9. (+9-6) & 2.489 & 0.273 & \cite{Garrett,HW,Guo76} \\
$^{74}$Se & 2.148 &  634.7 &  853.8 & 42.0 (6) & 77. (7)   & 1.345 & 1.833 & \cite{Garrett,Cottle42,McCutchan87,Budaca990} \\
$^{76}$Se & 2.380 &  559.1 & 1122.3 & 44. (1) & 47. (22)   & 2.007 & 1.068 & \cite{Garrett,HW,Budaca990} \\
$^{78}$Se & 2.449 &  613.7 & 1498.6 & 33.5 (8) & 1.17 (21) & 2.442 & 0.035 & \cite{Garrett,HW} \\
$^{80}$Se & 2.554 &  666.3 & 1478.8 & 24.7 (6) & 6.9 (11)  & 2.220 & 0.279 & \cite{Wood,HW} \\
$^{82}$Se & 2.650 &  654.8 & 1410.3 & 17.3 (10) & 3.62     & 2.154 & 0.209 & \cite{Wood,HW} \\

$^{74}$Kr & 2.225 &  455.6 &  509.0 & 67. (1) & 60. (17)     & 1.117 & 0.896 & \cite{Wood,HW,Garrett,Bender74,Clement75,Petrovici91,Nomura96} \\
$^{78}$Kr & 2.460 &  455.0 & 1017.2 & 67.9 (22) & 47. (4)    & 2.235 & 0.692 & \cite{HW,Garrett,Bender74,Nomura96} \\
$^{82}$Kr & 2.344 &  776.5 & 1487.6 & 21.3 (9) & 15. (5)     & 1.916 & 0.704 & \cite{HW,Nomura96} \\
$^{88}$Sr &       & 1836.1 & 3156.2 & 7.6 (4) & 4.0 (+15-14) & 1.719 & 0.526 & \cite{Maharana46,Xiang873} \\
$^{90}$Zr & 1.407 & 2186.3 & 1760.7 & 5.38 (13) & 26. (50)   & 0.805 & 4.833 &  \cite{HW,Maharana46,Xiang873,Togashi117} \\
$^{92}$Zr & 1.600 &  934.5 & 1382.8 & 6.4 (6) & 14.4 (5)     & 1.480 & 2.250 & 
 \cite{Wood,Garrett,Reinhard60,Xiang873,Togashi117,Gavrielov99,Gavrielov105} \\
$^{94}$Zr & 1.600 &  918.8 & 1300.2 & 4.9 (3) & 9.4 (4)      & 1.415 & 1.918 & 
 \cite{Wood,Garrett,Reinhard60,Ramos100,Ramos102,Xiang873,Togashi117,Nomura94,Gavrielov99,Gavrielov105} \\

$^{96}$Mo & 2.092 &  778.2 & 1148.1 & 20.7 (4) & 51. (7)        & 1.475 & 2.464 & \cite{Wood,Garrett,Abusara95,Xiang873,Nomura94} \\
$^{98}$Mo & 1.918 &  787.4 &  734.8 & 20.1 (4) & 48.5 (+50-125) & 0.933 & 2.413 & \cite{HW,Garrett,CDFTPRC,Thomas88,Abusara95,Xiang873,Nomura94} \\
$^{96}$\textbf{Ru} & 1.823 &  832.6 & 2148.8 & 18.4 (4) & 12. (+5-12)    & 2.581 & 0.652 & \cite{Abusara95} \\
$^{98}$\textbf{Ru} & 2.142 &  652.4 & 1322.1 & 29.8 (10) & 42. (+12-11)  & 2.026 & 1.409 &  \cite{Garrett,Abusara95,Nomura94} \\
$^{100}$\textbf{Ru}& 2.273 &  539.5 & 1130.3 & 35.7 (3) & 35. (5)        & 2.095 & 0.980 &  \cite{Garrett,Abusara95,Nomura94} \\
$^{102}$\textbf{Ru}& 2.329 &  475.1 &  943.7 & 44.6 (7) & 35. (6)        & 1.986 & 0.785 &  \cite{Garrett,Abusara95,Nomura94} \\

$^{104}$Pd& 2.381 &  551.8 & 1333.6 & 36.9 (19) & 13.2 (13) & 2.417 & 0.358 & \cite{Garrett} \\
$^{106}$Pd& 2.402 &  511.9 & 1133.8 & 44.3 (15) & 35. (8)   & 2.215 & 0.790 & \cite{Wood,HW,Garrett} \\
$^{108}$Pd& 2.416 &  433.9 & 1052.8 & 50.4 (15) & 52. (5)   & 2.426 & 1.032 & \cite{Wood,HW,Garrett} \\
$^{114}$Cd& 2.299 &  558.5 & 1285.6 & 31.1 (19) & 27.4 (17) & 2.032 & 0.881 & \cite{Wood,HW,Garrett,Nomura106} \\
$^{118}$Cd& 2.388 &  487.8 & 1134.5 & 33. (3) & 5.3 (8)     & 2.636 & 0.161 & \cite{Garrett} \\
$^{120}$Sn& 1.873 & 1171.3 & 1875.1 & 11.41 (22) & 12.6 (17)& 1.601 & 1.104 & \cite{Wood,Garrett} \\
 
$^{124}$Te& 2.072 &  602.7 & 1657.3 & 31.1 (5) & 20. (4)       & 2.750 & 0.643 & \cite{HW,Garrett,Sharma988} \\
$^{126}$Te& 2.043 &  666.4 & 1873.4 & 25.1 (5) & 8.8 (+8-11)   & 2.811 & 0.351 & \cite{HW,Garrett,Sharma988} \\
$^{144}$Nd& 1.887 &  696.6 & 2084.7 & 25.9 (5) & 19. (12)      & 2.993 & 0.734 & \cite{Wood} \\
$^{150}$Sm& 2.316 &  334.0 &  740.5 & 57.1 (13) & 53. (5)      & 2.217 & 0.928 & \cite{Wood,HW,Basak104} \\
$^{194}$Pt& 2.470 &  328.5 & 1267.2 & 49.5 (20) & 0.63 (+20-13)& 3.858 & 0.013 &  \cite{HW,Garrett,McCutchan71,Ramos84,Ramos89,Nomura83,Nomura84} \\
$^{198}$Pt& 2.419 &  407.2 &  914.5 & 31.81 (22) & 26. (7)     & 2.246 & 0.817 &  \cite{Garrett,Nomura83,Nomura84} \\

\hline

\end{tabular}
\end{table*}

\begin{table*}

\caption{Hg, Pb, and Po isotopes in which SC is known to occur, and for which the $0_2^+$ state is experimentally known \cite{ENSDF}.  Energies are given in keV, while $B(E2)$s are given in W.u.~. All data have been taken from Ref. \cite{ENSDF}. References discussing the presence of SC in the particular nucleus are given in the last column.  See Sec. II for further discussion. }

\bigskip

\begin{tabular}{ r r r r c r l  }

\hline
nuc. & $R_{4/2}$ & $E(2_1^+)$ & $E(0_2^+)$ & $B(E2; 2_1^+\to 0_1^+)$ & $R_{0/2}$ & Refs. \\
 \hline

$^{180}$Hg& 1.623 & 434.2 & 419.8 & 49 (9) & 0.967 & \cite{Dracoulis1988,Kondev2000,Elseviers2011,Yoshida1994,Patra1994,DeCoster1997,Nomura2013,Nomura2016b,GarciaRamos2014a,Delion2014,Jiao2015} \\
$^{182}$Hg& 3.198 & 351.7 & 328.0 & 55 (3) & 0.933 & 
\cite{Ma1984,Bree2014,WrzosekLipska2016,Yoshida1994,Patra1994,Richards1997,DeCoster1997,Shi2010,Nomura2013,Nomura2016b,GarciaRamos2014a,Delion2014,Jiao2015} \\
$^{184}$Hg& 2.962 & 366.8 & 375.1 & 62 (15)& 1.023 & 
\cite{Rud1973,Ma1986,Bree2014,Gaffney2014,WrzosekLipska2016,Bengtsson1987,Yoshida1994,Richards1997,DeCoster1997,Shi2010,Nomura2013,Nomura2016b,GarciaRamos2014a,Delion2014,Jiao2015} \\
$^{186}$Hg& 2.665 & 405.3 & 523.0 &71.3 (13)&1.290 & 
\cite{Proetel1973,Proetel1974,Hamilton1975,Cole1977,Ma1986,Bree2014,Gaffney2014,WrzosekLipska2016,Patra1994,Richards1997,DeCoster1997,Nomura2013,Nomura2016b,GarciaRamos2014a,Jiao2015} \\
$^{188}$Hg& 2.434 & 412.8 & 824.5 & 54 (9) & 1.997 & 
\cite{Hamilton1975,Cole1984,Bree2014,Gaffney2014,WrzosekLipska2016,Olaizola2019,Siciliano2020,Patra1994,DeCoster1997,Nomura2013,Nomura2016b,GarciaRamos2014a,Jiao2015} \\
$^{190}$Hg& 2.502 & 416.3 &1278.6 & 45 (3) & 3.071 & 
\cite{Kortelahti1991,Olaizola2019,Zhang1991,Nazarewicz1993,Patra1994,DeCoster1997,Nomura2013,Nomura2016b} \\

$^{184}$Pb&       &       & 570.0 &     &       & \cite{Cocks1998,Julin2016,Yoshida1994,Egido2004,RodriguezGuzman2004,Nomura2012b,Kim2022} \\
$^{186}$Pb&       &       & 530.0 &     &       &
\cite{Andreyev2000,Pakarinen2005,Julin2016,Heyde1989,Yoshida1994,Duguet2003,Frank2004,Egido2004,RodriguezGuzman2004,Hellemans2008,Shi2010,Nomura2012b,Jiao2015,Kim2022} \\
$^{188}$Pb& 1.470 & 723.6 & 591.0 &7 (3)& 0.817 & 
\cite{Allatt1998,LeCoz1999,Dracoulis2003,Dracoulis2004,WrzosekLipska2016,Heyde1989,Yoshida1994,Egido2004,RodriguezGuzman2004,Xu2007,Hellemans2008,Shi2010,Nomura2012b,Jiao2015,Kim2022} \\
$^{190}$Pb& 1.588 & 773.9 & 658.0 &     & 0.850 & 
\cite{WrzosekLipska2016,Heyde1989,Yoshida1994,Richards1997,Egido2004,RodriguezGuzman2004,Xu2007,Hellemans2008,Shi2010,Nomura2012b,Jiao2015,Kim2022} \\
$^{192}$Pb& 1.588 & 853.6 & 768.8 &     & 0.901 & 
\cite{VanDuppen1984,VanDuppen1987,WrzosekLipska2016,Heyde1989,Yoshida1994,DeCoster1997,Richards1997,Egido2004,RodriguezGuzman2004,Xu2007,Hellemans2008,Shi2010,Nomura2012b,Jiao2015,Kim2022} \\
$^{194}$Pb& 1.596 & 965.1 & 930.7 &     & 0.964 & 
\cite{VanDuppen1984,VanDuppen1987,WrzosekLipska2016,Heyde1989,Yoshida1994,DeCoster1997,Richards1997,Egido2004,Xu2007,Hellemans2008,Shi2010,Jiao2015,Kim2022} \\

$^{196}$Po& 1.924 & 463.1 & 558.0 &     & 1.205 & 
\cite{Alber1991,Younes1997,Kesteloot2015,WrzosekLipska2016,DeCoster1997,DeCoster1999a,Oros1999,Xu2007,Shi2010,GarciaRamos2015,Yao2013} \\
$^{198}$Po& 1.915 & 604.9 & 816.0 &     & 1.349 & 
\cite{Alber1991,Younes1997,Kesteloot2015,WrzosekLipska2016,DeCoster1997,DeCoster1999a,Oros1999,Xu2007,Shi2010,Delion2014,GarciaRamos2015,Yao2013}  \\
$^{200}$Po& 1.918 & 665.9 &1136.5 &     & 1.707 & 
\cite{Younes1997,Bijnens1998,Kesteloot2015,WrzosekLipska2016,DeCoster1999a,Oros1999,Shi2010,GarciaRamos2015,Yao2013} \\

\hline

\end{tabular}
\end{table*}

\begin{table}

\caption{Pb and Po isotopes in which predictions for  $B(E2;0_2^+\to 2_1^+)$ have been made (labeled by th) and/or data (label by exp) exist. $B(E2)$s are given in W.u.~. References  are given in the last column.  See Sec. II for further discussion. }

\bigskip

\begin{tabular}{ r r c c r l  }

\hline
nuc. & & $B(E2; 2_1^+\to 0_1^+)$ & $B(E2; 0_2^+\to 2_1^+)$ & $R_{0/2}$ & Refs. \\
 \hline
 
$^{186}$Pb& th & 21   &   30 &  1.429 & \cite{Nomura2012b} \\
$^{188}$Pb& th & 16   & 285  & 17.813 &  \cite{Nomura2012b} \\

$^{198}$Po& th & 39   &   7  & 0.179  & \cite{GarciaRamos2015} \\
$^{198}$Po& exp & 39 (9)   &   285 (+980-285)  & 7.308  & \cite{GarciaRamos2015} \\

\hline

\end{tabular}
\end{table}


\begin{figure*} [htb]

    {\includegraphics[width=75mm]{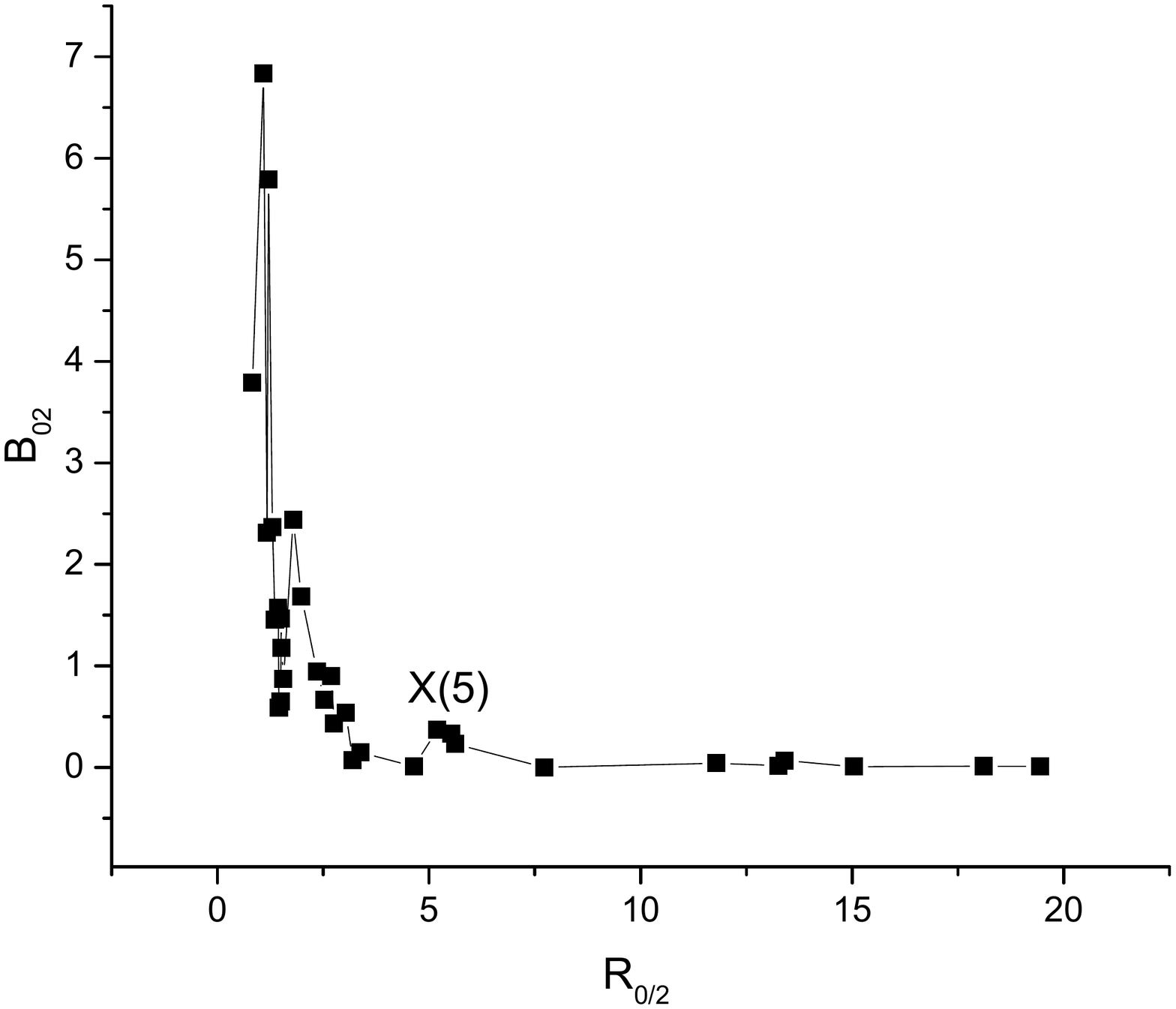}\hspace{5mm}
    \includegraphics[width=75mm]{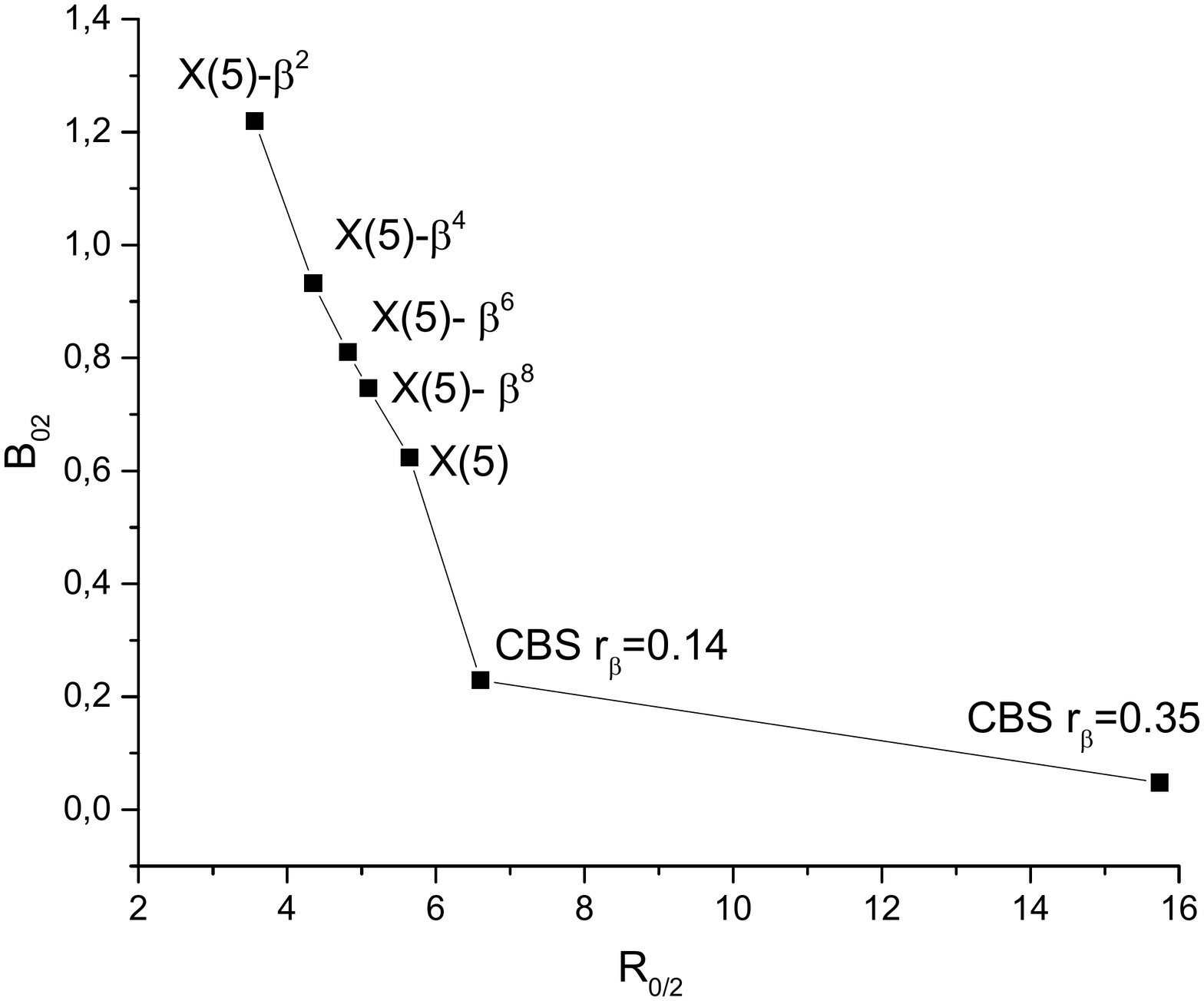}}

    \caption{(a) Ratios of transition rates $B_{02}$ (Eq. (\ref{B02})) are plotted vs. the energy ratios $R_{0/2}$ (Eq. (\ref{R02})) for all nuclei listed in Table I. The nuclei 
$^{40}$Ca and $^{98}$Zr, having ($B_{02}$, $R_{0/2}$) values (52.448, 0.639) and (50.0, 0.698) respectively, are not shown because of their very high values of $B_{02}$.    (b) Ratios of transition rates $B_{02}$ (Eq. (\ref{B02})) are plotted vs. the energy ratios $R_{0/2}$ (Eq. (\ref{R02})) for the X(5) critical point symmetry \cite{IacX5}, the X(5)-$\beta^{2n}$, $n=1,2,3,4$ models \cite{BonX5}, and two parameter values of the confined $\beta$-soft (CBS) model \cite{Gorbachenko}. In both panels lines have been drawn to guide the eye. The different scale in the vertical axes should be noticed.  See Sec. II for further discussion. }
    \label{map}
\end{figure*}

\begin{figure*}[htb]  

\includegraphics[width=160mm]{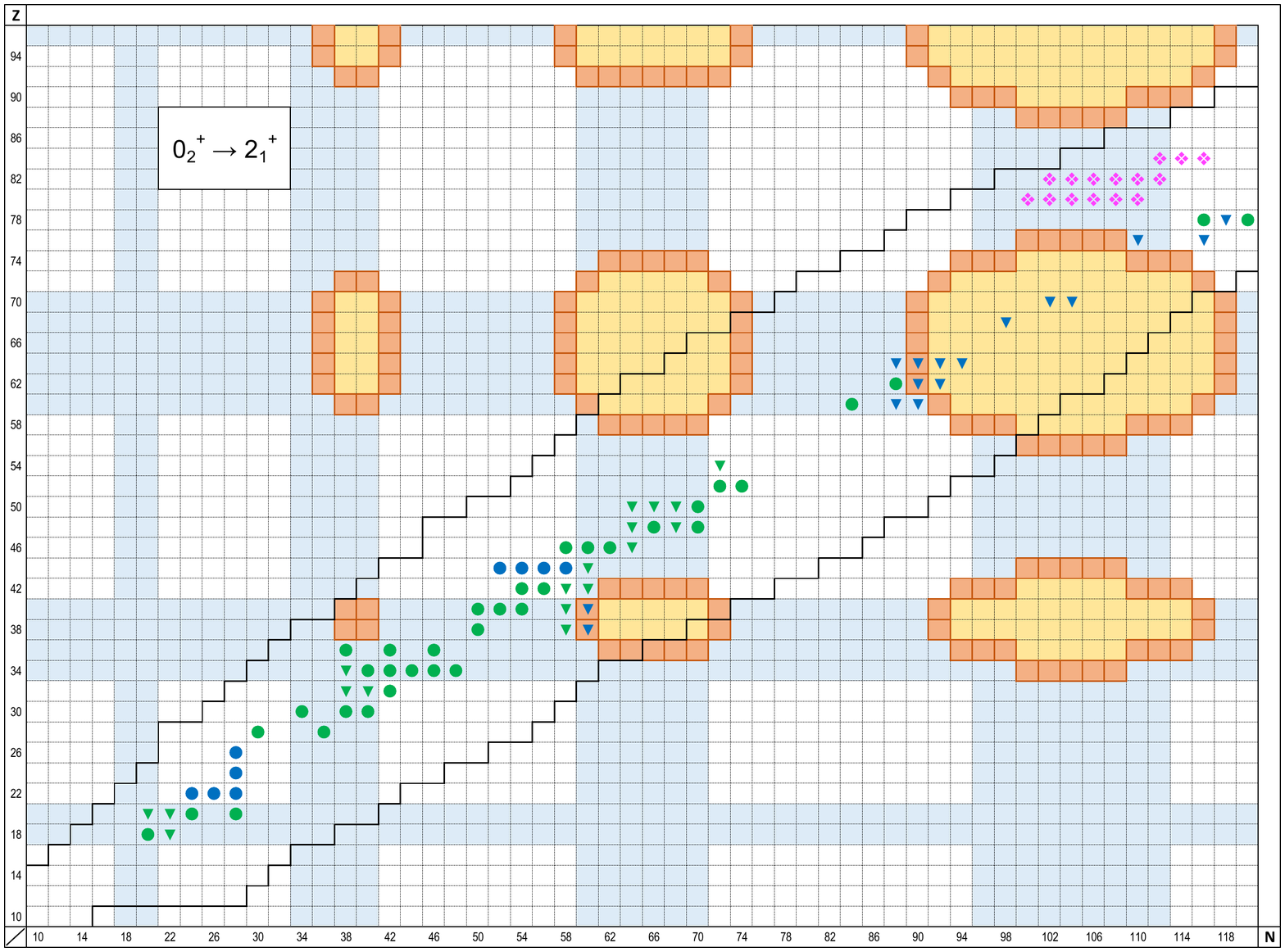}

\caption{
The nuclei with experimentally known \cite{ENSDF} $B(E2;0_2^+\to 2_1^+)$ transition rates and $K=0$ bands, listed in Table I, are shown on the nuclear chart. Nuclei of Table I  with $R_{4/2} <3.05$ are shown by  green full triangles, while deformed nuclei with $R_{4/2}\geq 3.05$ are shown by blue full triangles. In addition, the nuclei with experimentally known \cite{ENSDF} $B(E2;0_2^+\to 2_1^+)$ transition rates, but with no $K=0$ band based on the $0_2^+$ state known, listed in Table II, are shown. 
 Nuclei for which SC is known or expected to occur, according to the references given in Table II,  are shown as green full circles, while nuclei for which absence of SC is expected or predicted, according to the references given in Table II, are shown as blue full circles. Hg, Pg, and Po isotopes in which SC is known to occur and the $0_2^+$ state is known, appearing in Table III, are shown as purple diamonds. Furthermore, the azure stripes \cite{EPJASC} within which shape coexistence can be expected to occur according to the dual shell mechanism developed within the framework of the proxy-SU(3) symmetry \cite{proxy1,proxy2,EPJASM} are shown, along with  the contours  (indicated by orange boxes) corresponding to $P\approx 5$ \cite{McCutchan,CastenPPNP,Brenner}, within which (yellow areas) shape coexistence is not expected to occur according to the systematics of $B_{02}$ vs. $R_{0/2}$ discussed in the present work. The proton (neutron) drip lines, as extracted from the two-proton (two-neutron) separation energies taken from the database NuDat3.0 \cite{nudat3}, are shown as black thick lines. See Sec. II for further discussion. 
}  

\end{figure*}


\begin{figure*}[htb]  

{\includegraphics[width=75mm]{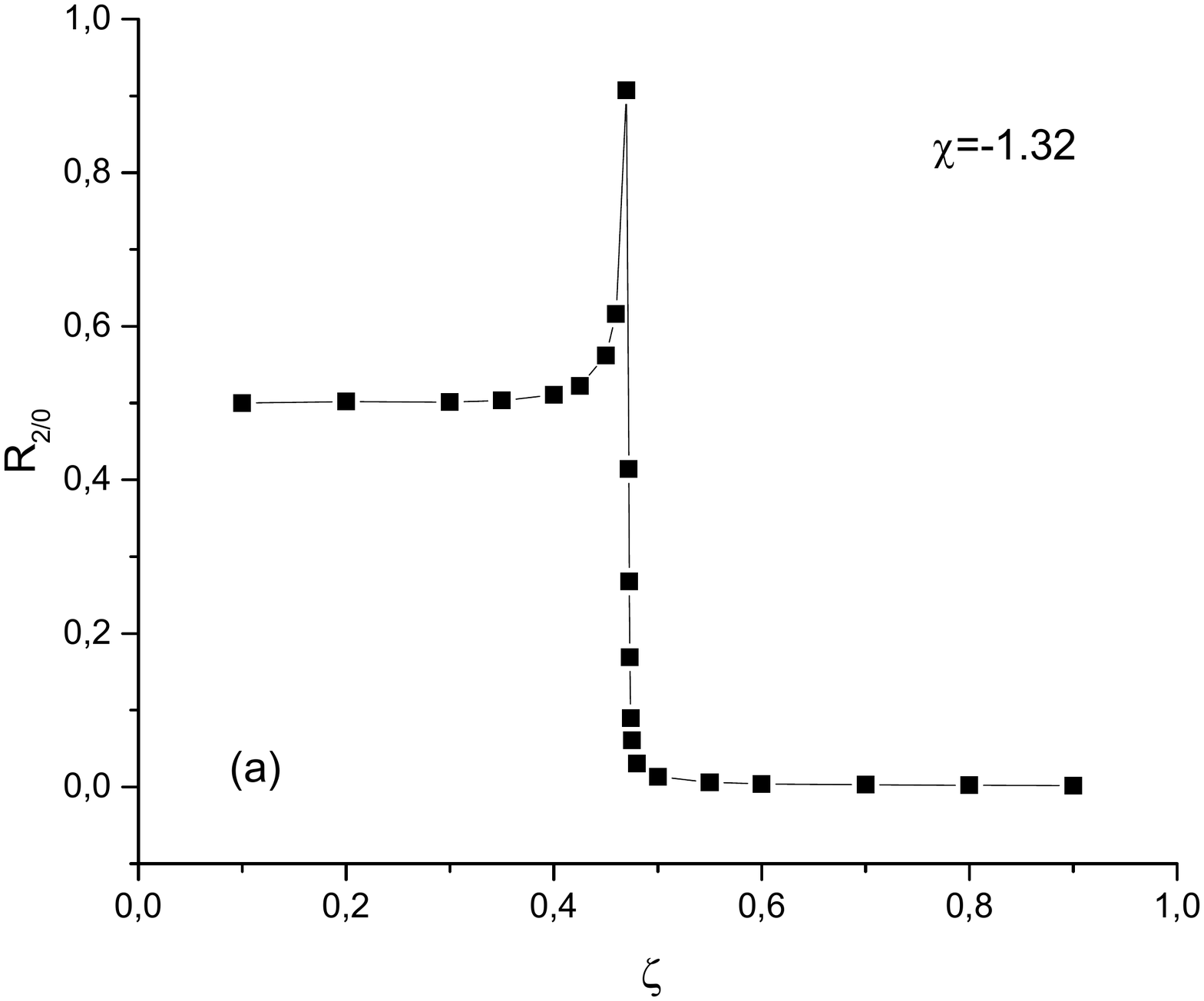}\hspace{5mm}
\includegraphics[width=75mm]{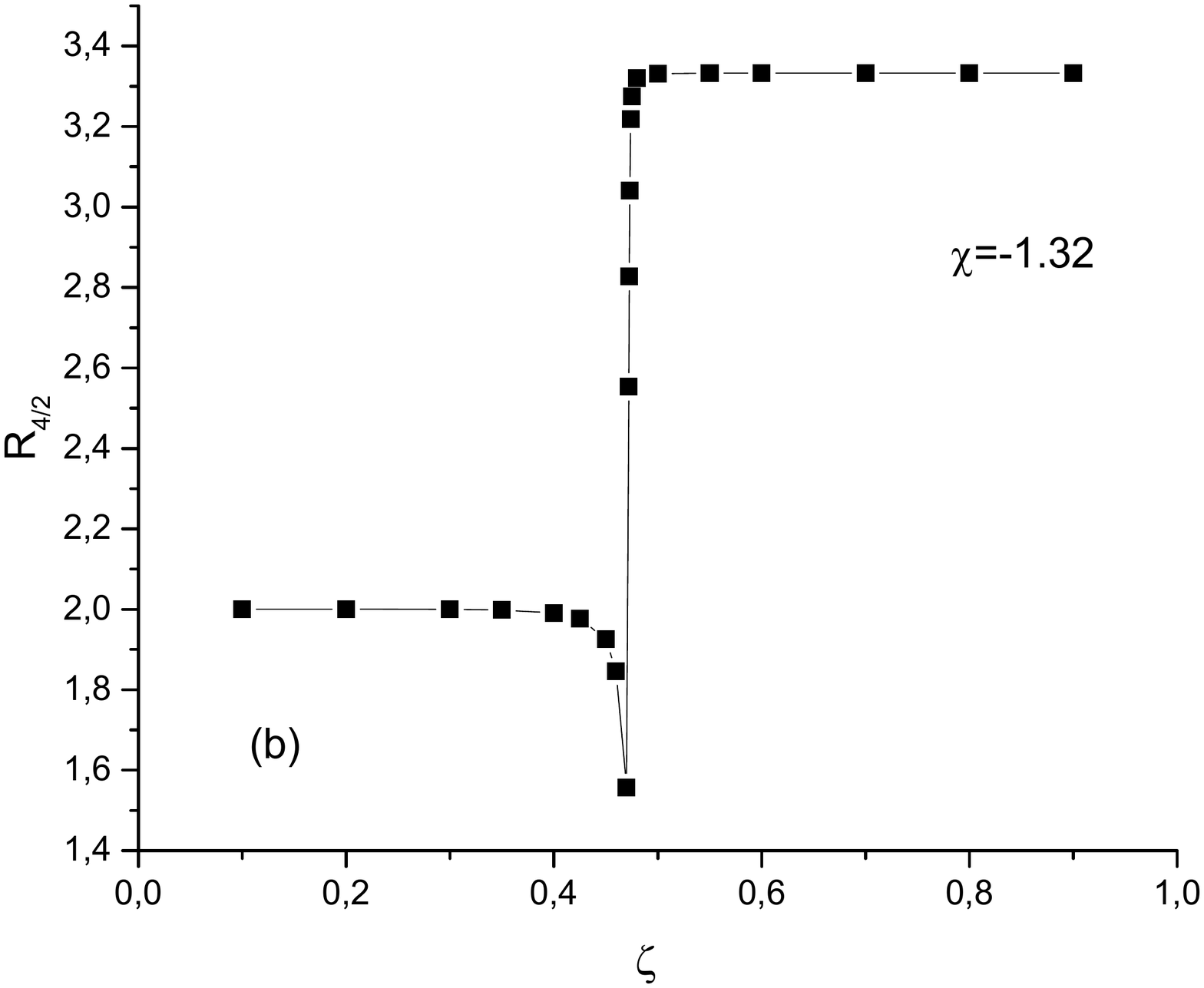}}

\caption{$R_{2/0}$ (Eq. \ref{R20})   and $R_{4/2}$ (Eq. \ref{R42}) energy ratios calculated with  the IBM Hamiltonian of Eq. (Eq. \ref{H})  with $\chi-\sqrt{7}/2$ and $N_B=250$, using the code IBAR \cite{IBAR}, are plotted vs. the control parameter $\zeta$. The critical point is located at $\zeta_{crit}=0.4721$. 
A large number of bosons has been used in order to exhibit the changes more clearly, since these are moderated at realistic boson numbers, as seen for example in Ref. \cite{order}. See Sec. III for further discussion.} 

\end{figure*}


\begin{figure}[htb]  

\includegraphics[width=75mm]{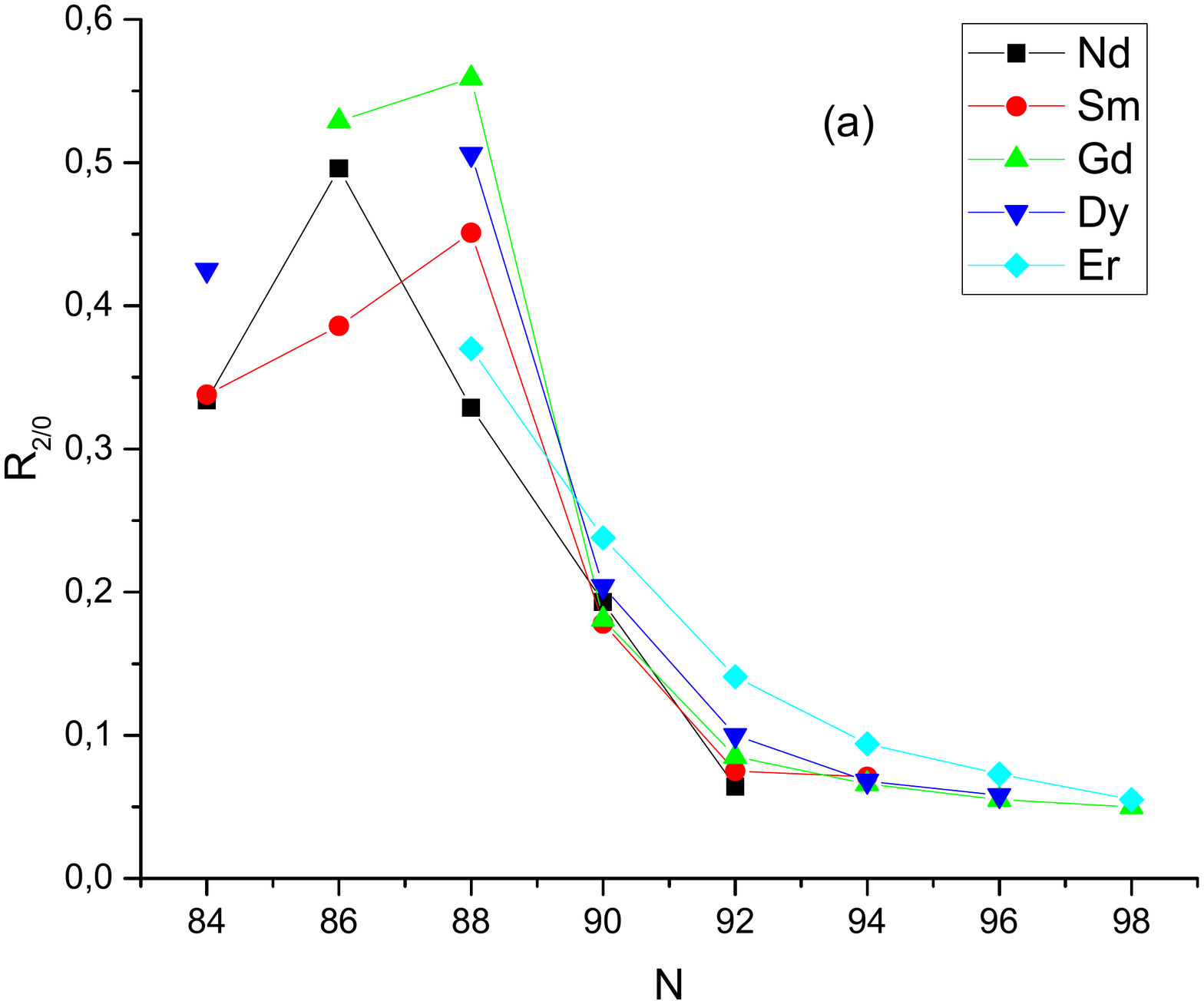}
\includegraphics[width=75mm]{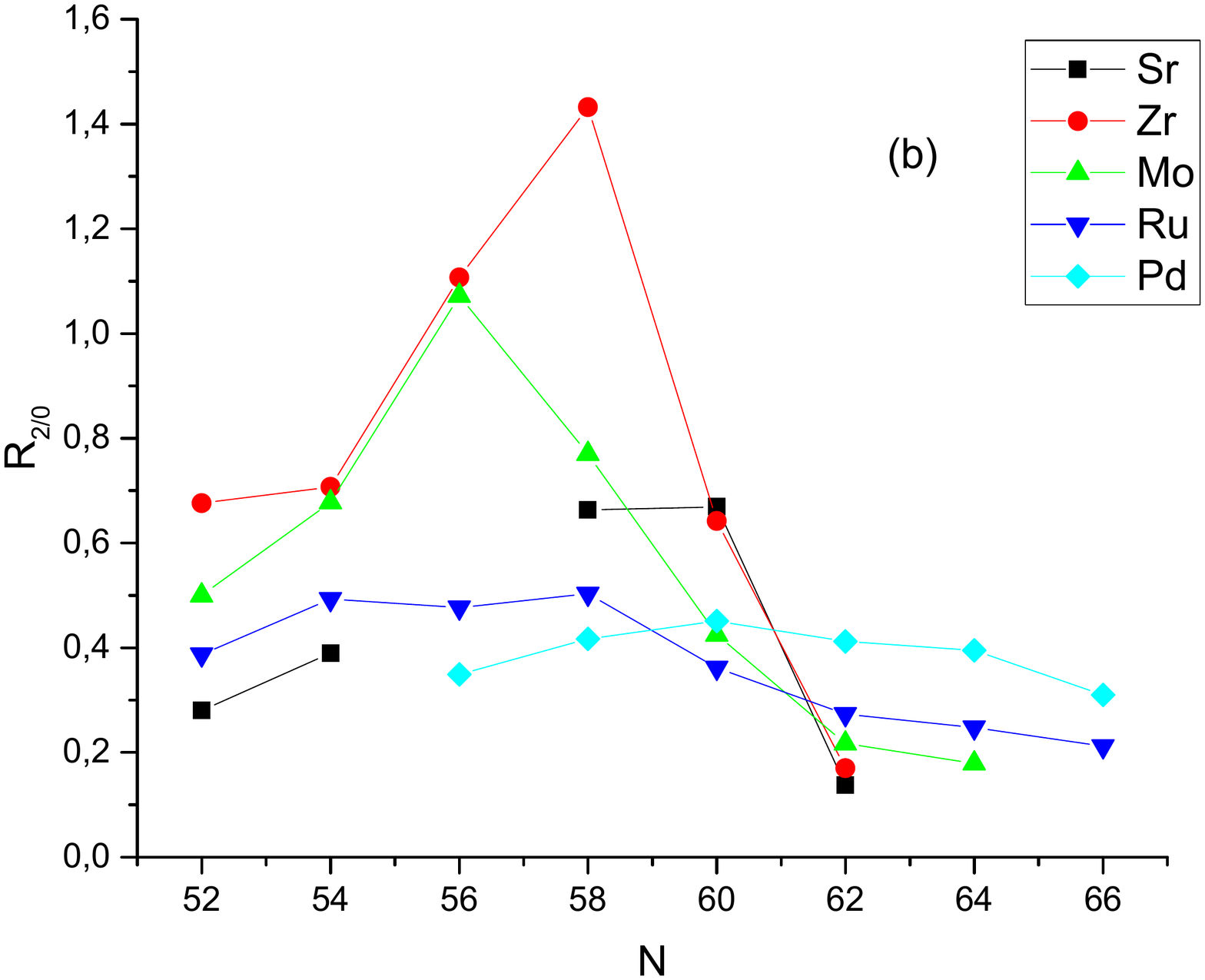}
\includegraphics[width=75mm]{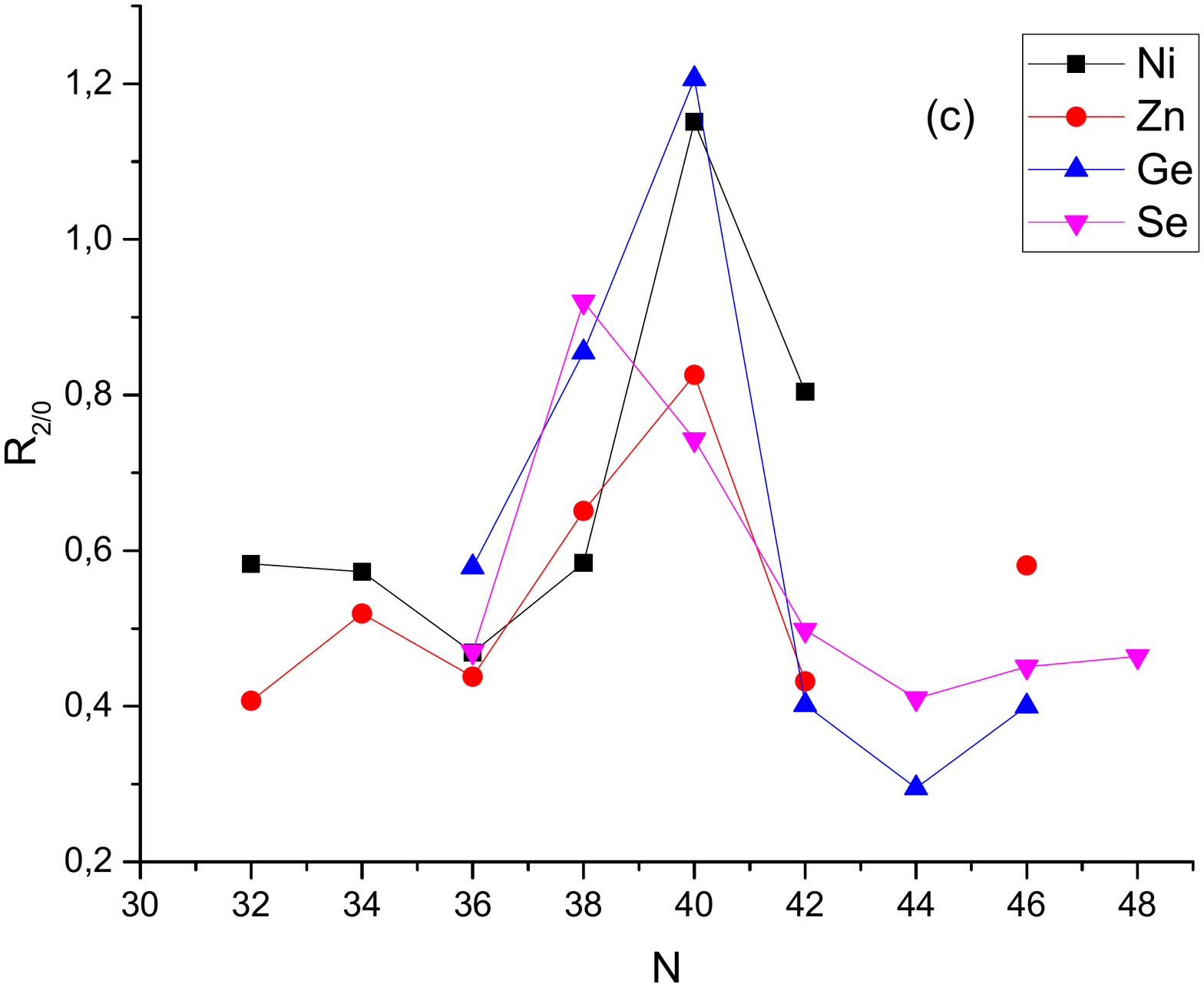}

\caption{Experimental $R_{2/0}$ energy ratios for various series of isotopes in different regions of the neutron number $N$, calculated from data taken from Ref. \cite{ENSDF}. The abrupt change of the ratio $R_{2/0}$ at $N=90$ (a), $N=60$ (b), and $N=40$ (c) is similar to the theoretical result obtained within the IBM, shown in Fig. 3(a), signifying the shape/phase transition from spherical to deformed shapes. The similarity among the three panels is due to the common mechanism behind these shape/phase transitions, due to the beginning of participation of the relevant neutron intruder orbital to the onset of deformation  \cite{FP1,FP2,FP3}. See Sec. III for further discussion.} 

\end{figure}

\end{document}